\RequirePackage{fix-cm}
\documentclass{svjour3}                     % onecolumn (standard format)
\smartqed  % flush right qed marks, e.g. at end of proof
\usepackage{graphicx}
\begin{document}
	\title{Study of autonomous conservative oscillator using an improved perturbation method%\thanks{Grants or other notes
		%about the article that should go on the front page should be
		%placed here. General acknowledgments should be placed at the end of the article.}
	}
	%\subtitle{Do you have a subtitle?\\ If so, write it here}
	
	%\titlerunning{Short form of title}        % if too long for running head
	
	\author{C. F. Sagar Zephania         \and Tapas Sil
		 %etc.
	}
	
	%\authorrunning{Short form of author list} % if too long for running head
	
	\institute{Department of Physics, Indian Institute of Information Technology Design and Manufacturing Kancheepuram, Chennai-600127, Tamil Nadu, India  \at
		%first address \\
		%Tel.: +123-45-678910\\
		%Fax: +123-45-678910\\
		%\email{fauthor@example.com}           %  \\
		%             \emph{Present address:} of F. Author  %  if needed
		\and
		Department of Physics, Indian Institute of Information Technology Design and Manufacturing Kancheepuram, Chennai-600127, Tamil Nadu, India  \at
		\email{tapassil@iiitdm.ac.in}
	}
	
	\date{Received: date / Accepted: date}
	% The correct dates will be entered by the editor

	\maketitle
	
%

%
%\offprints{\email{tapassil@iiitdm.ac.in}}          % Insert a name or remove this line
%
%\institute{Department of Physics, Indian Institute of Information Technology Design and Manufacturing Kancheepuram, Chennai-600127, Tamil Nadu, India 
%}
%
%\date{Received: date / Revised version: date}
% The correct dates will be entered by Springer
%
\abstract{
	 In a recent article \cite{manimegalai2019}, Aboodh transform based homotopy perturbation method ($ATHPM$) has been found to produce approximate analytical solutions in a simple way but with better accuracy in comparison to those obtained from some of the established approximation methods \cite{mehdipour2010application,nofal2013analytical} 
	 for some physically relevant anharmonic oscillators such as, autonomous conservative oscillator (ACO).  
	 In this article, expansion of frequency ($\omega$) and an auxiliary parameter ($h$) are introduced in the framework of homotopy perturbation method (HPM) to improve the accuracy by retaining its simplicity. Laplace transform is used to make the calculation simpler. This improved HPM ($LTHPMh$) is simple but provides  highly accurate results for ACO  in comparison to those from $ATHPM$.
%\keywords{Aboodh transform \and Homotopy Perturbation Method \and Anharmonic oscillator}
\PACS{31.15.xp, \and 43.40.Ga}
%\PACS{
%      {PACS-key}{discribing text of that key}   \and
%      {PACS-key}{discribing text of that key}
%     } % end of PACS codes
} %end of abstract
\section{Introduction}
\label{intro}
Autonomous conservative oscillator is an anharmonic oscillator with strong nonlinearity. 
Anharmonic oscillators are the systems generalizing the simple linear harmonic oscillator and are widely used for modeling many physical phenomena \cite{bonham1966use,bender1969anharmonic,chang1975quantum,hsue1984cs,ishmukhamedov2017tunneling,prentice2017first}. Finding solutions to the problems involving anharmonicity are difficult and most of them are not exactly solvable. Although, solving the anharmonic oscillator problems numerically are sometimes easy, one desires to get the analytical solutions of such problems as they carry more information and hence give a better insight about the system. There are many techniques for solving nonlinear oscillator problems such as the harmonic balance method \cite {nayfehnonlinear},
% Krylov-Bogoliubov-Mitropolsky method \cite{bogoli͡ubov1961asymptotic}
weighted linearization method \cite{agrwal1985weighted}, perturbation procedure for limit cycle analysis \cite{chen1991perturbation}, modified Lindstedt-Poincare method \cite{cheung1991modified}, Adomian decomposition method \cite{adomian1988review}
%, Nikiforov-Uvarov method \cite{nikiforov1988special,bera2013exact} 
and so on.  Most of these methods are not only somewhat plagued with the complexity of calculation but also fails to handle  problems with strong nonlinearity properly. Recently, a few techniques are proposed to obtain an analytical solution to the dynamical   systems with nonlinearity which are found to yield very good results without any higher-order approximation \cite{marinca2019,he2019}. One still strives for finding a technique which is simple but yields accurate solutions to the governing equation of the anharmonic oscillators. 

Liao \cite{liao1992proposed,liao2009theorem} proposed an analytical method, in 1992, known as the homotopy analysis method (HAM) which introduces an embedding parameter to construct a homotopy of the given system and then analyzes it by means of the Taylor formula. Subsequently, by means of the property of homotopy, one can transform a nonlinear problem into an infinite number of linear subproblems, irrespective of the fact that  the nonlinear problem contains small parameters or not. Therefore, unlike the perturbation method, this method does not depend on the values (smallness) of the perturbing  parameter. Moreover, HAM provides a  way to ensure the convergence of  solution series introducing an auxiliary parameter.
%Liao and Chwang \cite{liao1998application} applied HAM to solve some nonlinear problems and obtained excellent results.
On the other hand, 
J. H. He developed the homotopy perturbation method (HPM) for solving nonlinear problems with given  initial or boundary condition \cite{he1999homotopy,he2000coupling}. HPM is found to be very efficient in solving several problems with strong non-linearity in classical \cite{he1999homotopy,he2000coupling,biazar2011new} as well as quantum mechanical domain \cite{bera2012homotopy}. In this method, the solution is given in an infinite series usually converging to an accurate solution \cite{yildirim2009homotopy,biazar2015}. In fact, HPM is a special case of HAM \cite{he2004,liao2005} providing  simpler calculation in solving problems. Hence, sometimes it compromises with the accuracy. We have considered an expansion of frequency ($\omega$)  in the  framework of HPM \cite{bera2012homotopy} to improve the accuracy of calculation and also adopted an auxiliary parameter ($h$) \cite{liao2009theorem} to control the convergence.
%Inspired by HAM, we have adopted an auxiliary parameter ($h$) and considered an expansion of frequency ($\omega$)  in the  framework of HPM.    
Laplace  transform (LT) has been applied for making the calculation of solving differential  equations  further simple \cite{arfken}. The Laplace transform based HPM with  $h$ ($LTHPMh$) gives a simple way to achieve very high accuracy in solving nonlinear equations. 

$LTHPMh$ is used to get the displacement $(x)$ and $\omega$ for strongly nonlinear ACO.
We compare $LTHPMh$ results to those obtained from $ATHPM$ and  Hamiltonian approach technique ($HT$) \cite{hermann2014} where the  results from numerical calculations ($RK4$) are considered as the benchmark for checking the accuracy.\\
This paper is organized as follows. In section \ref{sec2}, we demonstrate briefly the formulation of $LTHPMh$. Application of $LTHPMh$ for studying an autonomous conservative oscillator has been shown in section  \ref{sec3}. Finally,  we provide a brief discussion and conclusions in section  \ref{sec:4}.
%%%%%%%
\section{Formalism}\label{sec2}
Let us consider a nonlinear inhomogeneous differential equation,
\begin{equation}\label{eq1}
G[{\ddot x},{\dot x},x,g]=0, 
\end{equation}
subjected to the initial conditions,
\begin{eqnarray}\label{eq3}
x(0)&=&a,\nonumber\\
{\dot x}(0)&=&0,
\end{eqnarray}
where, $\dot{x}$ denote the differentiation of $x$ with respect to the time ($t$), $G$ is the second order nonlinear differential operator and $g$ is the inhomogeneous term.
We can construct a homotopy of equation (\ref{eq1}) as given below,
\begin{equation}\label{eq2}
(1-p)U[x-x_g]+hp G[{\ddot x},{\dot x},x,g]=0,
\end{equation}
where,
\begin{equation}\label{eq4}
U[x]=\ddot{x}+\omega^2 x.
\end{equation}
Here, $p$ ${\in}$ [0,1] is an embedding parameter and $x_g=a \cos\omega t$ is the initial approximation of $ x(t)$ satisfying the conditions in equation (\ref{eq3}) with frequency $\omega$. For $p=0$, equation (\ref{eq2}) is  exactly solvable, whereas $p=1$ corresponds to the nonlinear problem for which we are trying to find out the solution. The convergence control parameter $h$ is incorporated in equation (\ref{eq2}). 
Using the transformation, 
$\tau=\omega t={\frac{t}{\sqrt{\Lambda}}}$,  in equation (\ref{eq2}), one can write,
\begin{equation}\label{eq5}
(1-p)( x''+ x- x_g''-x_g)+phM[ x^{''},x^{'},x,g,\Lambda]=0,
\end{equation}
with the initial conditions,
\begin{eqnarray}\label{eq6}
x(\tau=0)=a,\nonumber\\
x'(\tau=0)=0,
\end{eqnarray}
where, the prime denotes the differentiation with respect to $\tau$ and $M$ stands for the second order nonlinear differential operator after the transformation is taken.
Applying, LT on both sides of the equation (\ref{eq5}) and using the derivative property, $L[x^{n}(\tau)]=s^nL[x(\tau)]-s^{n-1}x(0)-s^{n-2}x^{'}(0)- ....-x^{n-1}(0)$, of the LT \cite{arfken} and then applying initial conditions, we get,  
\begin{eqnarray}\label{eq8}
(1-p)L[x-x_g]
&=&\frac{-ph}{s^2+1} L\left\lbrack M[ x^{''},x^{'}, x, g, \Lambda]\right\rbrack.
\end{eqnarray}
Using inverse LT on both sides of equation (\ref{eq8}), we get,
\begin{equation}\label{eq9}
x
=x_g+p(x-x_g)-L^{-1}\left\lbrace \frac{p h}{s^2+1} L[M[ x^{''},x^{'}, x, g, \Lambda]]\right\rbrace.
\end{equation}
Using HPM, we can expand  $x$ and $\Lambda$ in power series of $p$ as follows,
\begin{equation}\label{eq10}
x=\sum\limits_{n=0}^{\infty}x_{n}p^{n}=x_{0}+x_{1}p+x_{2}p^{2}+...,
\end{equation}
\begin{equation}\label{eq11}
\Lambda=\sum\limits_{n=0}^{\infty}\Lambda_{n}p^{n}=\Lambda_{0}+\Lambda_{1}p+\Lambda_{2}p^{2}+...
\end{equation}
Substituting equations (\ref{eq10}) and  (\ref{eq11}) in equation (\ref{eq9}) and equating the coefficients $p^0,p^1,p^2$..., we get the contributions  from higher order approximations to the displacement, such as,
\begin{eqnarray}\label{eq12-14}
p^0&:&\quad x_0(\tau)=x_g, \\
p^1&:&\quad	 x_1(\tau)=L^{-1}\left\lbrace\frac{-h}{s^2+1}\left(L[f_0[x^{''}_0, x^{'}_0, x_0 ,\Lambda_0]]\right)\right\rbrace, \\
p^2&:&\quad	x_2(\tau)
=x_1(\tau)-L^{-1}\left\lbrace \frac{h}{s^2+1}L[f_1[x^{''}_0, x^{'}_0, x_0,x_1'', x_1', x_1, \Lambda_0,  \Lambda_1]]\right\rbrace,
\end{eqnarray}
where, the $f_0$ and $f_1$ are  functionals.
The approximate solution is obtained putting $\tau=\omega t $ and $p= 1$,
\begin{equation}\label{eq15}
x(t)=x_0(t)+x_1(t)+x_2(t)+...,
\end{equation}
\begin{equation}\label{eq16}
\omega=(\Lambda_{0}+\Lambda_{1}+\Lambda_{2}+...)^{-1/2}.
\end{equation}
We calculate the average value of square residual of the differential equation for choosing the proper value of  $h$ \cite{liao2009theorem}, 
\begin{equation}\label{eq17}
E_{m}(h)=\frac{1}{2\pi}\int\limits_{0}^{2\pi}[\Delta_m(\tau,h)]^{2}d\tau,
\end{equation}
where,
\begin{equation}\label{eq18}
\Delta_{m}(\tau,h)=N[\tilde{x}^{''},\tilde{x}^{'}, \tilde{x},g,\tilde{\Lambda}],
\end{equation}
is the residual of the governing equation, and
\begin{equation}\label{eq19}
\tilde{x}=\sum\limits_{n=0}^{m}x_{n}(\tau),\quad \tilde{\Lambda}=\sum\limits_{n=0}^{m}\Lambda_{n}.
\end{equation}
In equation (\ref{eq19}), $\tilde{x}(\tau)$ and $\tilde{\Lambda}$,  corresponds to the approximation up to $m^{th}$ order of $x(\tau)$ and $\Lambda$, respectively. For converged solution, $E_m(h)$  should be minimum. Therefore,
\begin{equation}\label{eq20}
\frac{dE_m(h)}{dh}=0.
\end{equation}
The parameter $h$  can be chosen by evaluating the minimum value of the averaged square residual, $E_m(h)$, from equation (\ref{eq20}). 
For the purpose of computational efficiency, equation (\ref{eq17}) is discretized as,
\begin{equation}\label{eq21}
E_{m}(h)=\frac{1}{q+1}\sum\limits_{k=0}^{q}[\Delta_{m}(\tau_{k},h_0)]^2; \quad \tau_{k}=\frac{2k\pi}{q},
\end{equation}
where $q$ is an integer.  The value of $q$ is taken as $50$ for the calculations presented in the article.
\section{Applications}\label{sec3}
The equation of motion of the autonomous conservative oscillator (ACO) is given by,
\begin{equation}\label{eq22}
\ddot{x}(1+\epsilon x^{2}+\alpha x^{4}) +\lambda x+\epsilon x \dot{x}^{2}+ 2\alpha x^{3}\dot{x}^{2}+\beta x^{3}+\gamma x^{5}=0, 
\end{equation}
with initial conditions, $x(0)=a$ and $\dot{x}(0)=0$. The parameter $\lambda$ is an integer which may take values from $-1,0$ and $1$ whereas $\varepsilon, \alpha, \beta$ and $\gamma$ are positive parameters. In this article, $\lambda=1$ is considered for computation of displacements and frequencies of ACO for different values of the other parameters.
Homotopy of equation (\ref{eq22}) is constructed as follows,
\begin{equation}\label{eq23}
(1-p)U[x-x_g]+ hp\left(\ddot{x}(1+\epsilon x^{2}+\alpha x^{4})+\lambda x
+\epsilon x\dot{x}^{2}+2\alpha x^{3}\dot{x}^{2}+\beta x^{3}+\gamma x^{5}\right)=0.
\end{equation}
Introducing the transformation, $\tau=\frac{t}{\sqrt{\Lambda}}$ in equation (\ref{eq23}), we get,
\begin{eqnarray}\label{eq24}
(1-p)[ x''+ x- x_g''- x_g]+ph\left( x''(1+\epsilon x^{2}+\alpha x^{4})+\Lambda \lambda x 
+\epsilon x x'^{2}  \right.\nonumber\\
\left.+2 \alpha x^{3}x'^2+\Lambda \beta x^{3}+\Lambda \gamma x^{5}\right)=0,
\end{eqnarray}
where prime denotes the differentiation with respect to $\tau$.
Using LT on both sides of equation (\ref{eq24}) and using the derivative property of the LT \cite{arfken} and initial conditions followed by inverse $LT$,%\begin{eqnarray}\label{eq25}
%(1-p)L\left\lbrack x-x_g\right\rbrack =-\frac{ph}{(s^2+1)}L\left\lbrack x''(1+\epsilon x^{2}  +\alpha x^{4})+\Lambda\lambda x+\epsilon x x'^2\right.\nonumber \\
%\left.
%+2\alpha x^{3}x'^2+\Lambda\beta x^{3}+\Lambda\gamma x^{5}\right\rbrack.
%\end{eqnarray}
we may write,
\begin{eqnarray}\label{eq26}
x(\tau)= x_g +p(x-x_g)-hL^{-1} \left\lbrace \frac{1}{s^2+1}pL\left\lbrack x''(1+\epsilon x^{2} +\alpha x^{4})+\Lambda\lambda x \right.\right.\nonumber\\
\left.\left.+\epsilon xx'^2+2\alpha x^{3}x'^2+\Lambda\beta x^{3}+\Lambda\gamma x^{5} \frac{}{}\right\rbrack \right\rbrace.
\end{eqnarray}
Substituting equations (\ref{eq10}) and (\ref{eq11}) in (\ref{eq26}) and equating the coefficient of $p^0,p^1,p^2$.. etc., we get,
\begin{eqnarray}\label{eq27}
p^0&:&\quad x_0(\tau)=x_g,\\ 
\label{eq28}
p^1&:&\quad x_1(\tau)=-hL^{-1} \left\lbrace \frac{1}{s^2+1}L\left\lbrack x_0''(1+\epsilon x_0^{2}  +\alpha x_0^{4})+\Lambda_0\lambda x_0 +\epsilon x_0x_0'^2\right.\right.\nonumber\\
&&\quad \quad\quad \quad \left.\left.+ 2\alpha x_0^{3}x_0'^2+\Lambda_0\beta x_0^{3} 
+\Lambda_0\gamma x_0^{5}\right\rbrack\frac{}{}\right\rbrace,\\ 
\label{eq29}
p^2 &:&\quad x_2(\tau)= x_1  -hL^{-1}\left\lbrace\frac{1}{s^2+1}L \left \lbrack\frac{}{} \gamma \Lambda_1 x_0^{5}+x_1 \left\lbrace \lambda \Lambda_0+ \epsilon x_0'^{2}\right\rbrace+ x_0^3 \left\lbrace \beta\Lambda_1 \right.\right.\right.\nonumber\\
&&\quad \quad\quad \quad \left.+ 4\alpha x_0' x_1'+4\alpha x_1x_0''\right\rbrace +x_0 \left\lbrace \lambda \Lambda_1 +2\epsilon x_0'x_1' +2\epsilon x_1x_0''\right \rbrace +x_1'' \nonumber\\
&&\quad \quad\quad \quad\left.\left.+x_0^4\left \lbrace 5\gamma\Lambda_0 x_1+\alpha x_1''\right\rbrace+\left\lbrace \epsilon x_1''+3x_1\left(\beta \Lambda_0+2\alpha x_0'^2\right)\right\rbrace x_0^2 \frac{}{}\right \rbrack \right\rbrace,
\label{eq29a}
%p^3 &:&\quad x_3(\tau)= L^{-1}\left\lbrace \frac{1}{s^2+1} L[ x_2''(\tau)+ x_2(\tau)]\right\rbrace \nonumber \\
%&&  -hL^{-1}\left\lbrace\frac{1}{s^2+1}L[\Lambda _2 x_0(\tau) \left(\lambda +\beta  x_0{}^2(\tau)+\gamma  x_0{}^4(\tau)\right)+\Lambda _1\left(\lambda\right.\right.\nonumber\\
%&&\left.\left. +3 \beta  x_0{}^2(\tau)+5 \gamma  x_0{}^4(\tau)\right) x_1(\tau)+3 \beta  \Lambda _0 x_0(\tau) x_1{}^2(\tau)\right.\nonumber\\
%&&+10 \gamma  \Lambda _0 x_0{}^3(\tau) x_1{}^2(\tau)+\lambda
%\Lambda _0 x_2(\tau)+3 \beta  \Lambda _0 x_0{}^2(\tau) x_2(\tau)\nonumber\\
%&&+5 \gamma  \Lambda _0 x_0{}^4(\tau) x_2(\tau) +6 \alpha  x_0(\tau) x_1{}^2(\tau) x_0'{}^2(\tau)+\epsilon  x_2(\tau) x_0'{}^2(\tau)\nonumber\\
%&&+6 \alpha  x_0{}^2 (\tau)x_2(\tau) x_0'{}^2(\tau) +2 \epsilon  x_1(\tau) x_0'(\tau)
%x_1'(\tau) \nonumber\\
%&&+12 \alpha  x_0{}^2(\tau) x_1(\tau) x_0'(\tau) x_1'(\tau)+\epsilon  x_0(\tau) x_1'{}^2(\tau)+2 \alpha  x_0{}^3(\tau) x_1'{}^2(\tau)\nonumber\\
%&&+2 \epsilon  x_0(\tau) x_0'(\tau) x_2'(\tau)  +4 \alpha  x_0{}^3 (\tau)x_0'(\tau) x_2'(\tau)+\epsilon  x_1{}^2(\tau) x_0''(\tau)\nonumber\\
%&&+6 \alpha
%x_0{}^2(\tau) x_1{}^2(\tau) x_0''(\tau)+2 \epsilon  x_0(\tau) x_2(\tau) x_0''(\tau) +4 \alpha  x_0{}^3(\tau) x_2(\tau) x_0''(\tau)\nonumber\\
%&& +\left. 2 \epsilon
%x_0(\tau) x_1(\tau) x_1''(\tau)+4 \alpha  x_0{}^3(\tau) x_1(\tau) x_1''(\tau)+x_2''(\tau)+\epsilon  x_0{}^2 (\tau)x_2''(\tau)\right. \nonumber\\
%&& \left.+\alpha  x_0{}^4(\tau) x_2''(\tau)]\frac{}{}\right\rbrace.
\end{eqnarray}
Using the properties  of LT followed by inverse LT in equation (\ref{eq28}), we get,
\begin{eqnarray}\label{eq32}
x_1(\tau)&=& \frac{h}{16}[8a+4\epsilon a^3+3 \alpha a^5-\Lambda_0(8a\lambda +6 \beta a^3+5\gamma a^5) ]\tau \sin\tau +\frac{h}{128}[ 8\epsilon a^3+7 \alpha a^5\nonumber\\
&&-4\Lambda_0 \beta a^3-5\gamma\Lambda_0 a^5] (\cos\tau-\cos 3\tau) +\frac{ha^5}{384}\left\lbrack 3 \alpha-\gamma\Lambda_0\right\rbrack (\cos\tau-\cos 5\tau). 
\end{eqnarray}
The coefficient   of $\tau\sin\tau$ in equation (\ref{eq32}) needs to be equal to zero for avoiding the secular term, i.e.,
\begin{equation}\label{eq33}
8a+4\epsilon a^3+3 \alpha a^5-\Lambda_0(8a\lambda +6 \beta a^3+5\gamma a^5) =0,
\end{equation} 
which gives $\Lambda_0$ as,
\begin{equation}\label{eq34}
\Lambda_0 =\frac{8+3a^4\alpha+4a^2\epsilon}{8 \lambda+5a^4\gamma+6a^2\beta}.
\end{equation}
Therefore, the frequency from the zeroth order  approximation is,
\begin{equation}\label{eq35}
\omega_0=\sqrt{\frac{8\lambda+5a^4\gamma+6a^2\beta}{8+3a^4\alpha+4a^2\epsilon}},
\end{equation}
which is same as the frequency from the first order approximation in ATHPM \cite{manimegalai2019} and HT \cite{hermann2014}. 
The remaining part of $x_1(\tau)$ in equation (\ref{eq32}) gives the first order approximation of $x$,
\begin{eqnarray}\label{eq36}
x_1(\tau)&=&\frac{h}{128}\left\lbrack 8\epsilon a^3+7 \alpha a^5-4\Lambda_0 \beta a^3-5\gamma\Lambda_0 a^5\right\rbrack (\cos\tau-\cos3\tau)+\frac{ha^5}{384}\left\lbrack 3 \alpha-\gamma\Lambda_0\right\rbrack\nonumber\\
&&\times (\cos\tau-\cos5\tau). 
\end{eqnarray}
Similarly, the coefficient of $p^{2}$ will also have a secular term and  must be zero to have a physical solution which gives the first order correction to $\Lambda$, as given below,
\begin{eqnarray}\label{eq38}
\Lambda_1&=&\frac{a^2 h}{192(6a^2\beta +5a^4\gamma+8 \lambda)}\left \lbrack 96a^2 \alpha -15a^6 \alpha^2 +96\epsilon-12a^4\alpha\epsilon
-\Lambda_0(48\beta+64a^2\gamma \right.\nonumber\\
&&+138 a^4\alpha\beta+150a^6\alpha\gamma+144a^2\beta\epsilon
+156a^4\gamma\epsilon+96a^2\alpha\lambda+96\epsilon\lambda)+\Lambda_0^2(72a^2\beta^2\nonumber\\
&&\left.+174a^4\beta\gamma +105a^6\gamma^2+48\beta\lambda+64a^2\gamma\lambda)\right \rbrack.
\end{eqnarray}
Therefore, considering first order correction to $\Lambda$, the frequency can be written as,
\begin{eqnarray}\label{eq39}
\omega=(\Lambda_0+\Lambda_1)^{-1/2}.
\end{eqnarray}
The displacement of ACO, considering of the term upto first order is obtained as,
\begin{eqnarray}\label{xl1}
x_{L}(t)&=& a\cos\omega t + \frac{h}{128}\left\lbrack 8\epsilon a^3+7 \alpha a^5-4\Lambda_0 \beta a^3-5\gamma\Lambda_0 a^5\right\rbrack (\cos\omega t-\cos3\omega t)
\nonumber\\
&&+\frac{ha^5}{384}[3 \alpha-\gamma\Lambda_0]\left(\cos\omega t-\cos5\omega t\right),
\end{eqnarray}
where, the frequency $\omega$ is  given in equation (\ref{eq39}).
%The analytical solution of eq.(\ref{eq22}) using $ATHPM$ considering the first order approximation is written as \cite{manimegalai2019},
%\begin{eqnarray}\label{xathpm} 
%x_{A}(t)&=& a\cos\omega {t} 
%+\frac{1}{128\omega^2}\left[8\epsilon a^3\omega^2+7\alpha a^5\omega^2-4\beta a^3 -5\gamma a^5\right] (cos{\omega}t\nonumber \\
%&&-cos3{\omega}t)+\frac{a^5}{384\omega^2}\left[3\alpha\omega^2\right.\left. -\gamma\right] (cos{\omega}t-cos5{\omega}t),
%\end{eqnarray}
%with frequency $\omega$, which  is the same as obtained from $LTHPMh$ for the zeroth apprximation given in eq.(\ref{eq35}).
%%%%
\begin{figure}
	\centering
	\resizebox{0.49\textwidth}{!}{\includegraphics{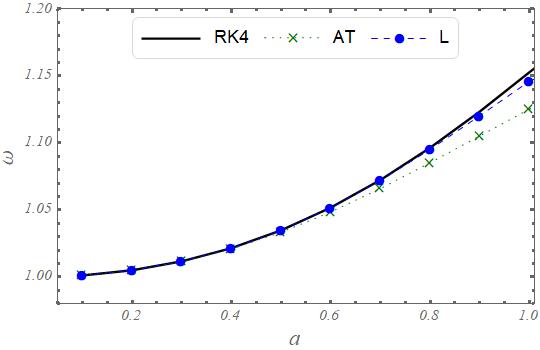}}
	%	\vspace{2cm}       % Give the correct figure height in cm
	\caption{The frequency ($\omega$) of ACO obtained from ATHPM and LTHPMh are compared with those extracted from $RK4$ results for different values of $a$ with $OP=1.0$} 
	\label{freq}   % Give a unique label
\end{figure}
%%%%%  
\begin{figure}
	\centering
	\resizebox{0.49\textwidth}{!}{\includegraphics{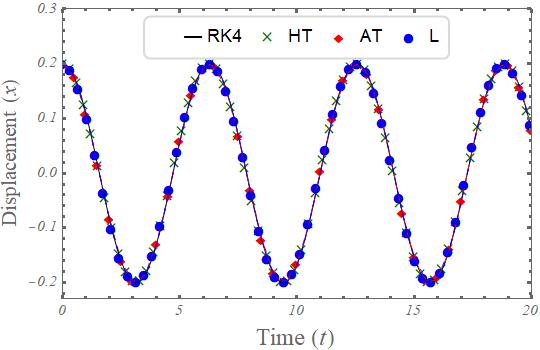}}
	\resizebox{0.49\textwidth}{!}{\includegraphics{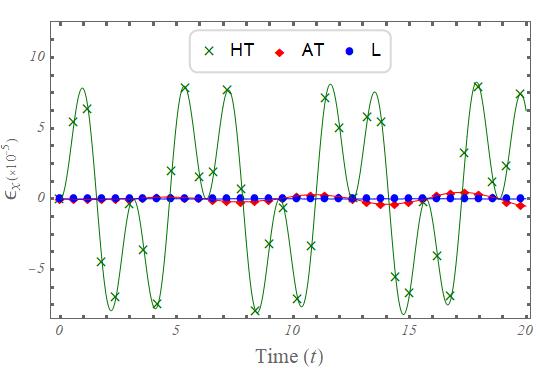}}
	\resizebox{0.49\textwidth}{!}{\includegraphics{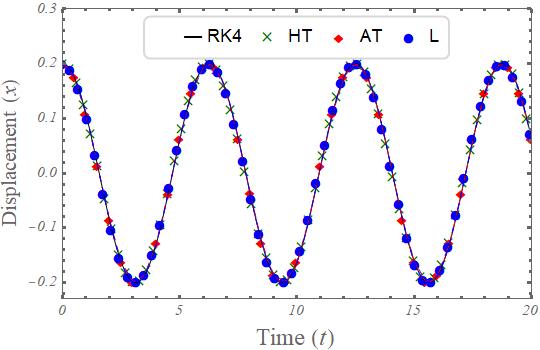}}
	\resizebox{0.49\textwidth}{!}{\includegraphics{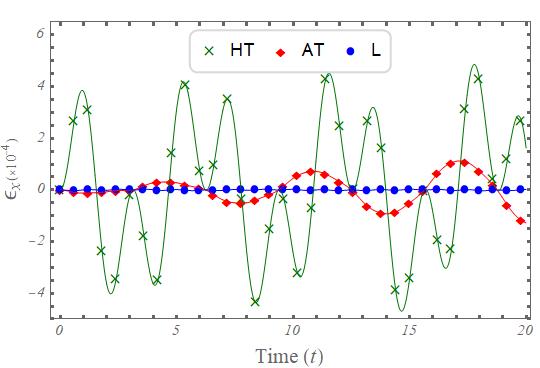}}
	\resizebox{0.49\textwidth}{!}{\includegraphics{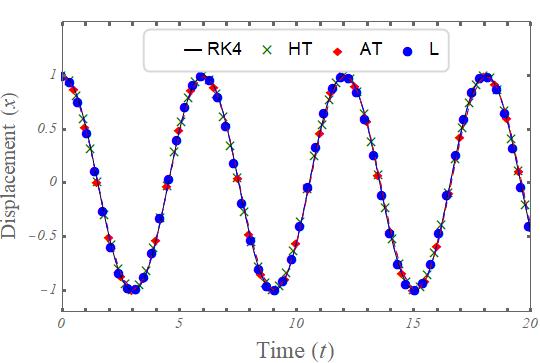}}
	\resizebox{0.49\textwidth}{!}{\includegraphics{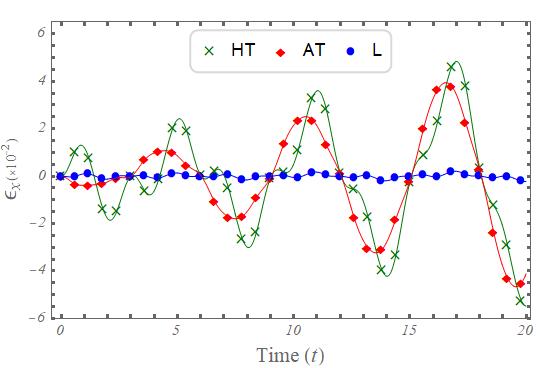}}
	\resizebox{0.49\textwidth}{!}{\includegraphics{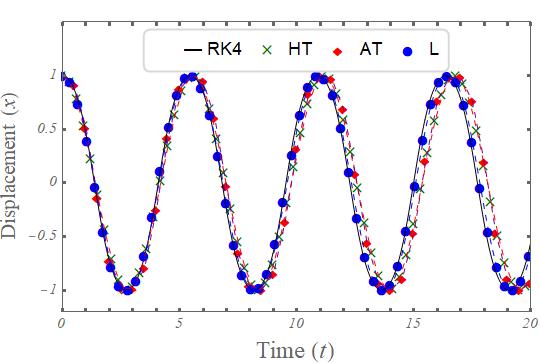}}
	\resizebox{0.49\textwidth}{!}{\includegraphics{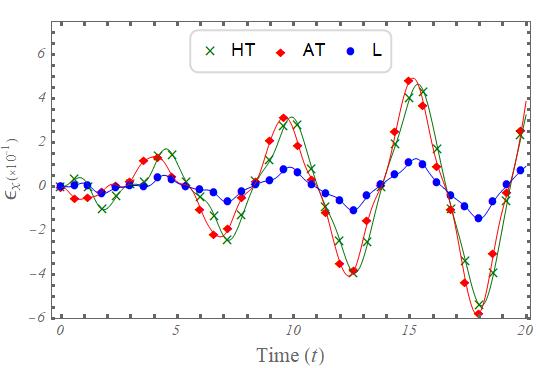}}
	%\vspace{2cm}       % Give the correct figure height in cm
	\caption{Plot of variation of displacements $x_{RK4}$ (black solid line), $x_{HT}$   (green cross),  $x_{AT}$   (red diamond), and  $x_{L}$ (blue circle) with time ($t$) are shown in the four panels of the left column for four parameter sets A, B, C, and D  starting from the  top panel  to bottom panel, respectively. Absolute errors in the approximate solutions ($HT$, $ATHPM$ and $LTHPMh$) with respect to $RK4$ for the same parameter sets are displayed in the right column.}
	\label{fig1}       % Give a unique label
\end{figure}
%%

%In this article, we have considered 6 different parameter sets, A($a=0.2,OP=0.2$), B($0.2,1.0$), C($1.0,0.2$), D($1.0,1.0$), E($1.0,0.5$) and F($0.5,1.0$), to discuss the autonomous conservative oscillator. 
We display the calculated values of frequency ($\omega$)
as a function of amplitude ($a$) with $OP=1$ using $ATHPM$ ($HT$), $LTHPMh$ and $RK4$ in figure \ref{freq}.
Here, $OP$ stands for the ``other parameters" given in equation (\ref{eq22}), {\it i.e.},  $OP=\alpha= \beta=\epsilon=\gamma$. It is seen from the figure that all approximate methods give same values as $RK4$   at the lower amplitude for example, $\omega_{APPROX}=1.00126$ at $a=0.1$. At a larger amplitude,  $LTHPMh$ frequency remain  very close to those obtained from $RK4$ whereas $ATHPM$ underestimates the $\omega$. The percentage error in $\omega$  are $2.385$ and $0.584$ for $ATHPM$ and $LTHPMh$, respectively, for $a=1$ which reduce to $0.167$ if the second order term in $LTHPMh$ is considered.

We plot the displacements obtained from $HT$ ($x_{HT}$, green cross) \cite{hermann2014} , $ATHPM$ ($x_{AT}$, red diamond) \cite{manimegalai2019} and  $LTHPMh$ ($x_{L}$, blue circles) expressed  in equation (\ref{xl1}) respectively, with  increasing $t$ for four sets  of parameters A($a=0.2,OP=0.2$), B($0.2,1.0$), C($1.0,0.2$) and D($1.0,1.0$) in the left column of figure~\ref{fig1}, and compared with the same obtained by numerical solution of  equation (\ref{eq22}) employing the fourth order Runge-Kutta ($RK4$) method  ($x_{RK4}(t)$,   black solid line).  It is seen that for small values of the parameters, approximate displacements match well with $x_{RK4})$ but a significant deviations for $x_{AT}$ and $x_{HT}$ from $x_{RK4}$ are noticed for large values of force parameters. The deviation (error) of the approximate  displacements  with respect to its values calculated using $RK4$ ($\epsilon_{x_{APPROX}}=x_{RK4}-x_{APPROX}$), $\epsilon_{x_L}$,  $\epsilon_{x_{AT}}$  and  $\epsilon_{x_{HT}}$  are displayed in the panels of right column of figure~\ref{fig1} for the same parameter sets as taken for calculating the graphs presented in the  corresponding panels of the left column.
%$ATHPM$ has been  shown to yield  better accuracy in calculating displacements for some anharmonic oscillators including ACO \cite{manimegalai2019} in comparison to well established FAF-EBM method \cite{nofal2013analytical}.
All panels in the right column show that accuracy of $x_{L}(t)$ is much improved  in comparison to  $x_{AT}(t)$ and $x_{HT}(t)$ throughout the range of time considered here.
%\begin{table}
%	\begin{center}
%		\caption{Values of $h$,  maximum deviation of approximate $x$ from $x_{RK4}$ and corresponding rms error for different sets of parameters. Here, $u(-n)$ indicates $u\times10^{-n}$. }\label{tab1}
%{\tiny
%		\begin{tabular}{|c|c|c|c|c|c|c|}
%			\hline
%			$a$ & OP & h & $\epsilon^{max}_{xA}$ & $\epsilon^{max}_{xL}$ & 
%			$\epsilon^{rms}_{xA}$ & $\epsilon^{rms}_{xL}$\\
%			%0.5,5.0,q
%			\hline
%			0.2 & 0.2 &0.996672 &-5.1016(-06) &-1.0949(-07) &2.1452(-06) & 6.0990(-08) \\
%			\hline
%			0.2&0.8 &  0.986653 &-8.1660(-05)  &-1.7657(-06)  & 3.3845(-05) &  9.4916(-07)\\
%			\hline
%			0.2 &  1.0 &0.983304  &   -1.2752(-04)  & -2.7601(-06)&5.2641(-05) & 1.4724(-06)\\
%			\hline
%			0.8 &	0.2 &0.903987 & -1.1605(-02) &2.6277(-04)  & 4.7546(-03) &  1.2546(-04)\\
%			\hline
%			0.8 &	0.8 & 0.667982 & -1.3142(-01)  & -1.3364(-02) & 5.4939(-02) &  4.8592(-03)\\
%			\hline
%			0.8	& 1.0 & 0.607391   & -1.8615(-01) & -2.4465(-02) & 7.6473(-02)&  8.9997(-03)\\  
%			\hline
%			1.0 &	0.2 & 0.818679 & -4.6617(-02) & -2.2540(-03)  & 1.9669(-02)&  8.0067(-04)\\
%			\hline
%			1.0 & 0.8 & 0.478981  & -4.3095(-01) & -9.0714(-02)  &  1.7129(-01) & 3.4287(-02)\\
%			\hline
%			1.0 &	1.0 &0.413602 &-5.8133(-01) & -1.4702(-01) &2.2853(-01) &  5.6200(-02)\\
%			\hline
%		\end{tabular}
%	\end{center}
%\end{table}
\begin{table}
	\begin{center}
		\caption{Values of $h$,  rms error of approximate $x$ with respect to $x_{RK4}$ for different sets of parameters. Here, $u(-n)$ indicates $u\times10^{-n}$. }\label{tab1}
		%{\tiny
		\begin{tabular}{|c|c|c|c|c|c|}
			\hline
			$a$ & OP & h &  $\epsilon^{rms}_{x_{HT}}$ & 
			$\epsilon^{rms}_{x_{AT}}$ & $\epsilon^{rms}_{x_L}$\\
			%0.5,5.0,q
			\hline
			0.2 & 0.2 &0.996672  &5.1200(-05) &2.1452(-06) & 6.0990(-08) \\
			\hline
			0.2&0.8 &  0.986653   &2.0514(-04)  & 3.3845(-05) &  9.4916(-07)\\
			\hline
			0.2 &  1.0 &0.983304    &2.5725(-04)&5.2641(-05) & 1.4724(-06)\\
			\hline
			0.8 &	0.2 &0.903987  &6.1287(-03)  & 4.7546(-03) &  1.2546(-04)\\
			\hline
			0.8 &	0.8 & 0.667982  &5.6830(-02) & 5.4939(-02) &  4.8592(-03)\\
			\hline
			0.8	& 1.0 & 0.607391    & 7.7976(-02) & 7.6473(-02)&  8.9997(-03)\\  
			\hline
			1.0 &	0.2 & 0.818679 &2.1314(-02)  & 1.9669(-02)&  8.0067(-04)\\
			\hline
			1.0 & 0.8 & 0.478981   &  1.6981(-01)  &  1.7129(-01) & 3.4287(-02)\\
			\hline
			1.0 &	1.0 &0.413602  &2.2266(-01) &2.2853(-01) &  5.6200(-02)\\
			\hline
		\end{tabular}
	\end{center}
\end{table}

\begin{figure}
	\centering
	%\resizebox{0.45\textwidth}{!}{\includegraphics{a10R08secnd_xt.jpg}}
	\resizebox{0.49\textwidth}{!}{\includegraphics{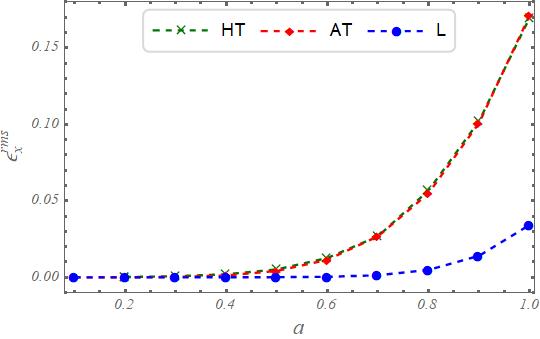}}
	%	\vspace{2cm}       % Give the correct figure height in cm
	%\caption{Plot of $\epsilon_{x}^{rms}$  with increasing $a$ for different values of $OP$.}
	%\label{fig6}   % Give a unique label
	%\end{figure}
	%%
	%%
	%\begin{figure}
	%	\centering
	\resizebox{0.49\textwidth}{!}{\includegraphics{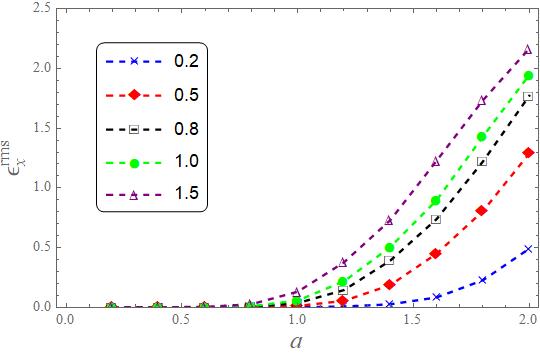}}
	%	\resizebox{0.45\textwidth}{!}{\includegraphics{h_variation.jpg}}
	\caption{ Comparison  of $\epsilon_{x}^{rms}$(rms of $\epsilon_{x}$)   form $HT$, $ATHPM$ with  $LTHPMh$  approximation with $OP=0.8.$ is displayed in the left panel. Plot of $\epsilon_{x}^{rms}$  with increasing $a$ for different values of $OP$ (right panel).
	}\label{fig4}       
\end{figure}

In order to get a clear idea about the performance of $LTHPMh$, we display
in table~\ref{tab1},  the root mean square deviations ($\epsilon^{rms}_{x_{APPROX}}$) of $x_{APPROX}$ from $x_{RK4}$   are presented in fourth, fifth and sixth columns, respectively. The root mean square deviation is  defined as,
\begin{equation}\label{rmsErr}
\epsilon^{rms}_{x_{APPROX}}(t)=\sqrt{\frac{1}{N}\sum\limits_{i=1}^{N}[x_{RK4}(t_i)-x_{APPROX}(t_i)]^2},
\end{equation}
where $N$ represent the maximum number of points considered.
Values of $h$  obtained from equation (\ref{eq20}) which are  used to calculate $x_{L}$ for different values of '$a$' (first column) and '$OP$' (second column) are tabulated in the third column. It is seen that $h$ is decreasing with the increase in the values of $a$ and/or $OP$. It is noticed from the table that both $\epsilon^{rms}_{x_{HT}}$ and  $\epsilon^{rms}_{x_{AT}}$ are order of magnitude higher in comparison to  $\epsilon^{rms}_{x_L}$. 
Variations of $rms$ errors in $x$ obtained  from $HT$ (green cross), $ATHPM$ (red diamond) and $LTHPMh$ (blue circle) 
% and $LTHPMh-2$ (green square) 
with  $a$ are presented in left panel of figure~\ref{fig4} for $OP=0.8$. It displays that the all curves almost coincide with each other (approximately zero error) for a low amplitude. The curves corresponding to $HT$ and $ATHPM$  start diverging rapidly  together from $LTHPMh$ curves approximately at   $a=0.4$ and reach the value approximately $0.17$ at $a=1$ whereas the rms error  of $x_L$ is $0.035$.
Therefore, $LTHPMh$ is found to yield  much improved results in comparison to those obtained from $HT$ and $ATHPM$.
Plot of $\epsilon_{x_L}^{rms}$  with increasing  amplitude for different values of $OP$ starting from $0.2$ to $1.5$ are presented in the right panel of figure~\ref{fig4}. The values of $\epsilon_{x_L}^{rms}$ remain very small up to $a=1.0$ for all values of $OP$ considered. The $rms$ error starts increasing beyond $a=1$ and reaches a  high value  at $a=2.0$. It is noted that the rate of  increase in $\epsilon_{x_L}^{rms}$ is more for a larger $OP$, i.e., stronger nonlinearity.
%The $rms$ error in $x$ from first order $LTHPMh$ starts slowly deviating from zero error line  and attain the values $\epsilon_{xL}^{rms}=0.034287$, at the highest amplitude considered here.

%%
\begin{figure}
	\centering
	\resizebox{0.49\textwidth}{!}{\includegraphics{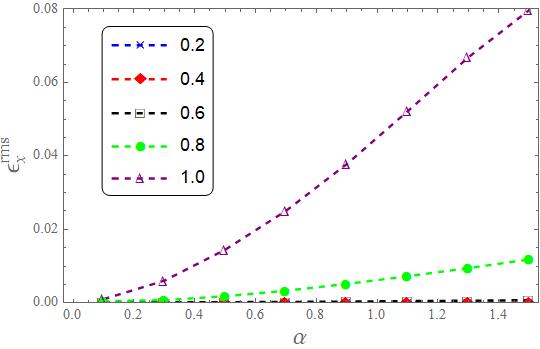}}
	\resizebox{0.49\textwidth}{!}{\includegraphics{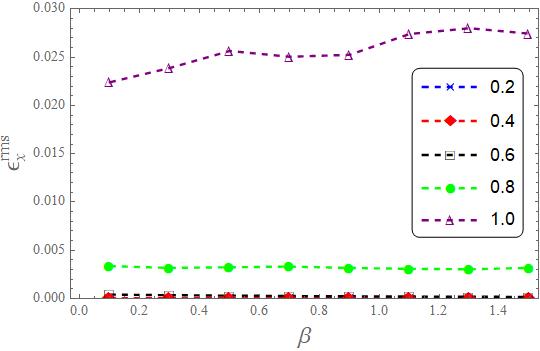}}
	\resizebox{0.49\textwidth}{!}{\includegraphics{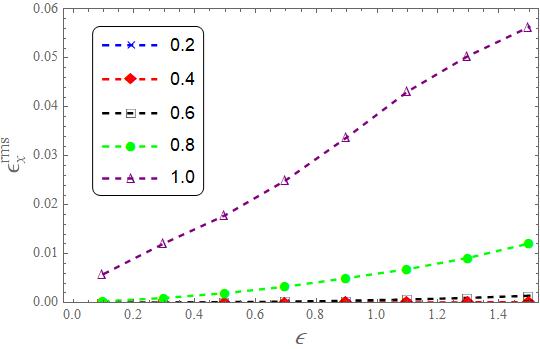}}
	\resizebox{0.49\textwidth}{!}{\includegraphics{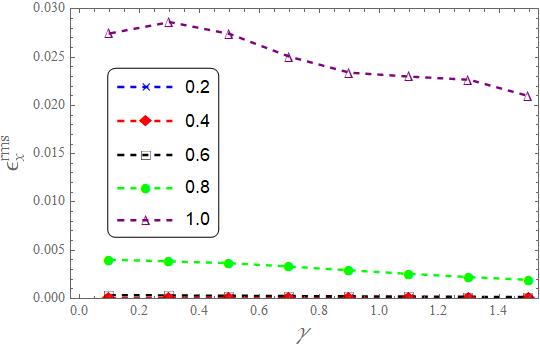}}
	%	\vspace{2cm}       % Give the correct figure height in cm
	\caption{Plot of $\epsilon_{x}^{rms}$  in $LTHPMh$ displacement with respect to the $\alpha$ (left top), $\beta$ (right top), $\epsilon$ (left bottom) and $\gamma$ (right bottom) for differnent amplitudes $a$ and $OP=0.7$.} 
	\label{fig5}   % Give a unique label
\end{figure}
\begin{figure}
	\centering
	\resizebox{0.49\textwidth}{!}{\includegraphics{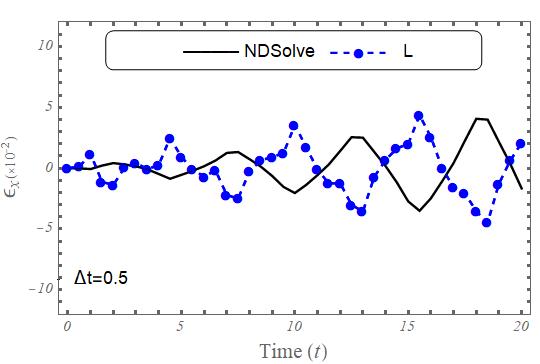}}
	\resizebox{0.49\textwidth}{!}{\includegraphics{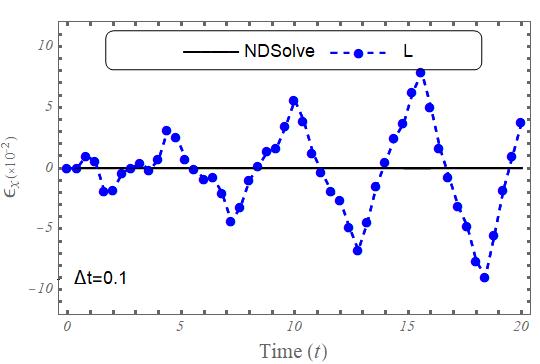}}
	\resizebox{0.49\textwidth}{!}{\includegraphics{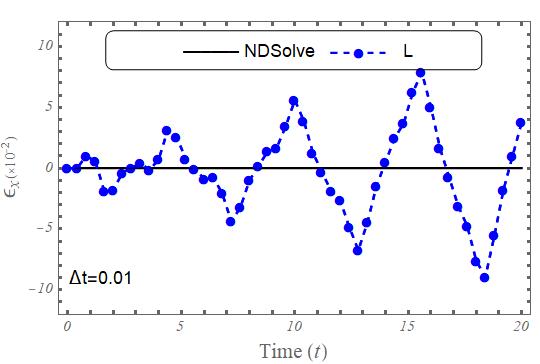}}
	\resizebox{0.49\textwidth}{!}{\includegraphics{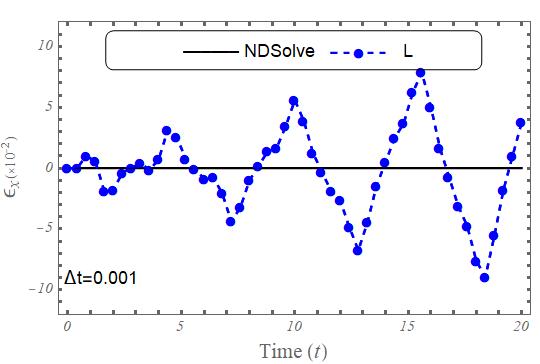}}
	\caption{Plot of  error in  displacement  obtained from Mathematica function $NDSolve$  ($\epsilon_{x_{NDS}}$, black dotted line), and $LTHPMh$ ($\epsilon_{x_L}$, blue circles) 
		with respect to those calculated from $RK4$ versus time for different values of the mesh size  ($\Delta t$) of time ($t$) for $a=1.0$ and $OP=0.8$.}
	\label{fig2}   
\end{figure}
\begin{figure}
	\centering
	\resizebox{0.49\textwidth}{!}{\includegraphics{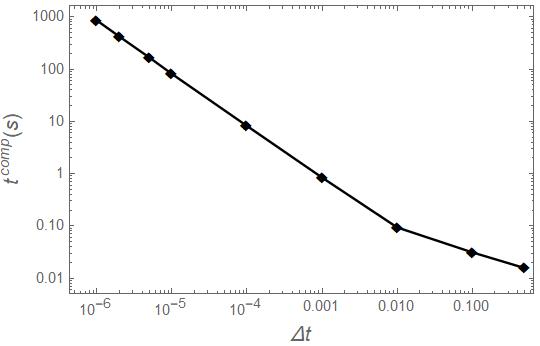}}
	%	\vspace{2cm}       % Give the correct figure height in cm
	\caption{Variation of computation time ($t^{comp}$) of $RK4$ calculation with step size in $t$ for $a=1$ and $OP=0.8$} 
	\label{RK4tMesh}   % Give a unique label
\end{figure}

The changes in  root mean square deviations of $x$ obtained from $LTHPMh$ with respect to the $RK4$ values ($\epsilon_{x_L}^{rms}$) with increasing $\alpha$ (left top), $\beta$ (right top), $\epsilon$ (left bottom) and $\gamma$ (right bottom) for different amplitudes ($a=0.2-1.0$ in steps of $0.2$) taking $OP=0.7$ are displayed in figure~\ref{fig5}. It is observed from the  left top panel that  $\epsilon_{x_L}^{rms}$ are very small at very low amplitudes for the entire range of $\alpha$ considered. The $rms$
errors increases  for large values of $\alpha$. The increase in $\epsilon_{x_L}^{rms}$ is more rapid for a larger value of $a$. The same is noted in the left bottom panel for $\epsilon$. Right two panels show that  $\epsilon_{x_L}^{rms}$  is not very sensitive to the variation of $\beta $ and $\gamma$ for a particular value of $a$. Though, the $rms$ error increases for the increase in $a$ for these two cases also.

\begin{figure}
	\centering
	\resizebox{0.45\textwidth}{!}{\includegraphics{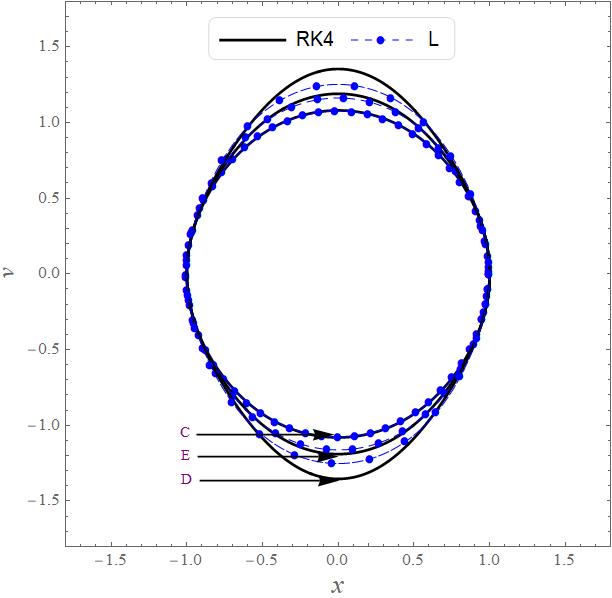}}
	\resizebox{0.45\textwidth}{!}{\includegraphics{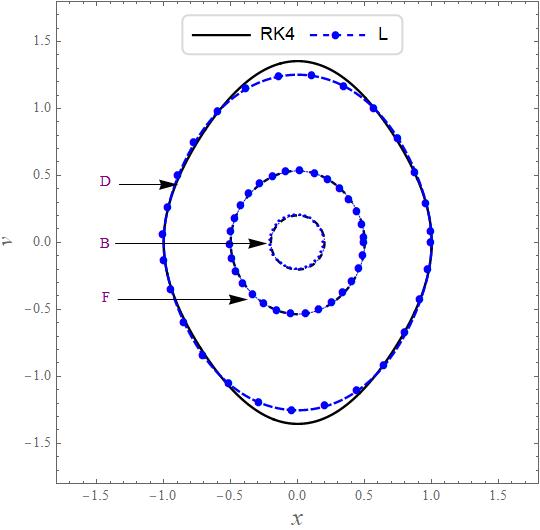}}
	\resizebox{0.45\textwidth}{!}{\includegraphics{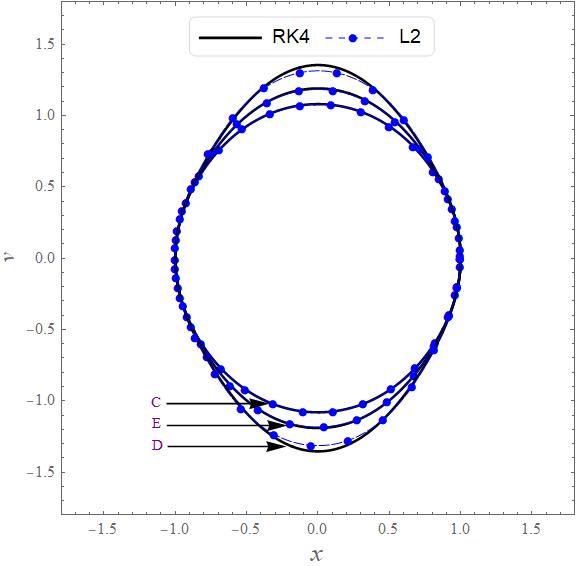}}
	\resizebox{0.45\textwidth}{!}{\includegraphics{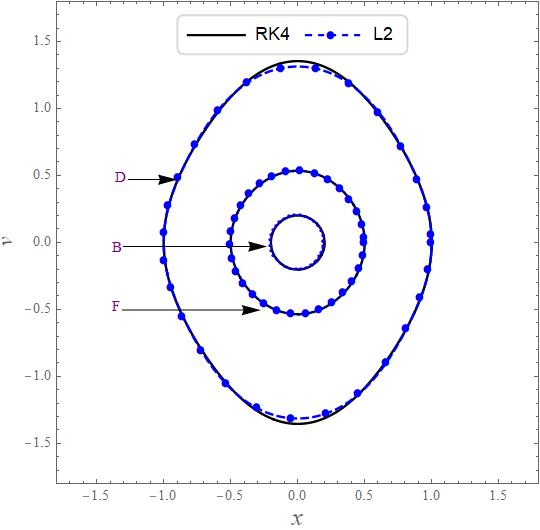}}	
	%	\resizebox{0.45\textwidth}{!}{\includegraphics{a10R10secnd_err.jpg}	
	%	\vspace{2cm}       % Give the correct figure height in cm
	\caption{Comparison of phase-space curve of the equation (\ref{eq22}) obtained from LTHPMh and RK4. Top left panel compare the phase space curve for different $OP$ for $a=1.0$ whereas the top right panel compare phase curve for different  $a$ for $OP=1.0$. Bottom two panels correspond to the second order $LTHPMh$ ($L2$) calculations}
	\label{phase}
	%\label{fig3}   % Give a unique label
\end{figure}
%\newpage

Comparisons of the results obtained from $LTHPMh$, $ATHPM$ and $HT$ are presented in  previous figures and table considering the  $RK4$ results as the reference. In order to check the reliability of $RK4$ results, we compute errors in  displacement  obtained from Mathematica function $NDSolve$ ($\epsilon_{x_{NDS}}$, black solid line) and $LTHPMh$ ($\epsilon_{x_L}$, blue circles) 
with respect to those calculated from $RK4$. The variations of  $\epsilon_{x_{NDS}}$ and $\epsilon_{x_{L}}$ with $t$ are displayed in figure~\ref{fig2} for different  mesh sizes  ($\Delta t = 0.5, 0.1, 0.01$ and $0.001$).
It is observed that $\epsilon_{x_{NDS}}$ remains  very close to zero throughout the span of time considered here for all values of $\Delta t$ except $0.5$. Moreover,  the profile of the $\epsilon_{x_L}$ curves and their maximum values  are  matching extremely well for all values of $\Delta t$ except for $0.5$.  This ascertains the accuracy of the numerical solutions using $RK4$ if $\Delta t<0.1$. 
Computation times ($t^{comp}$) for the solution of equation (\ref{eq22}) with different $\Delta t$ for $a=1.0$ and $OP=0.8$ is plotted in figure~\ref{RK4tMesh}
(Mathematica is used in a computer with i7-7700, 3.60GHz and RAM=16GB). It is noted that $t^{comp}$ starts increasing by an order of magnitude for the decrease in $\Delta t$ by ten times from $\Delta t=0.01$ onwards. 
Therefore, the mesh size, $\Delta t = 0.001$ is taken as the optimum one considering the convergence of solution  and  computation time. We have used this mesh size  for all calculation presented in this article. It is noted that computational time for the calculation of $x$ and $\omega$ using $ATHPM$ is $7.35938s$ whereas the time increases to $13.5625s$
and $59.5313s$ for the first order and second order $LTHPMh$ methods, respectively (for $a=1$ and $OP=1$).      

Significant information about the motion and stability of a system can be extracted from phase portrait analysis. We plot the velocity, $v(=dx/dt)$ versus $x$ obtained from $LTHPMh$ and $RK4$ for the system described in equation  (\ref{eq22}). In the top left panel of the figure \ref{phase}, the phase-space curves are displayed for the parameter set $C$, $D$ and  $E($1.0,0.5$)$ which shows that the approximate phase-space curve ($LTHPMh$)  matches extremely well with $RK4$ curve for lower value of $OP$
for  amplitude $a=1$ but for the higher values of the $OP$ , i.e.  when the nonlinearity becomes stronger, the $LTHPM$ phase curves deviate  from the circular one and becomes more prolate keeping the center at origin $(0,0)$. In the top right panel of figure \ref{phase}, phase-space curves are presented for different values of $a$ with $OP=1.0$ (parameter sets $B$, $D$ and $F($0.5,1.0$)$). Similar deviations of $LTHPMh$ curves from $RK4$  is observed for larger nonlinearity of the system due to the  increased amplitudes. Both the panels of figure \ref{phase} confirms the periodic nature of the system with single stability point at the origin  $(0,0)$ for the parameter sets considered here. The second order $LTHPMh$ phase curves ($L2$) are compared with $RK4$ results  in bottom two panels and are found to exhibit similar trends but with better accuracy.  Therefore, $LTHPMh$ can be reliably used  for the phase-curves analysis.
%%%%%
%%%%%
\section{Conclusion}
\label{sec:4}
An improved HPM ($LTHPMh$) is introduced to study oscillators with strong anharmonicity such as ACO. 
HPM is a simpler method than the HAM but in some cases it fails to yield accurate solution. In this article, a convergence parameter is introduced and  the expansion of frequency  term is also considered in the framework of HPM . Laplace transform is used to make the calculation easy. 
$LTHPMh$ is used to find the approximate analytical expression of displacement and  frequency  of oscillation for strongly nonlinear ACO and
are compared to those obtained from $HT$, $ATHPM$ and  numerical calculations ($RK4$). It is found that the new method gives the values of  displacement and  frequency  with an accuracy at least one order of  magnitude better than  those of  $ATHPM$ and $HT$ for the parameter sets considered here. Comparison of $rms$ deviations of displacement obtained from three aforementioned approximate methods are in  corroborate with the conclusion that $LTHPM$ is a better approximate method.
Results on $rms$ deviations of displacements obtained from $LTHPMh$ quantify the effect of the strength of nonlinearity on the accuracy of the approximate result.    Contributions of higher order terms are found to be nontrivial in calculating displacement and frequency of ACO, vibrating for long time and/or with large amplitude. $LTHPMh$ has been found to be trustworthy for analyzing phase portrait of a system.
Computations of displacement and frequency for the cases considered in this article are very fast. 
%It is noted that $LTHPMh$ yields better accuracy in comparison to $ATHPM$ and which has been found to be more efficient and accurate than some established approximate methods.
It is to conclude that $LTHPMh$ is a blend of simplicity  and delicacy. Hence, it can be used to study physical problems with strong nonlinearity, efficiently.


\begin{thebibliography}{}
%
\bibitem{manimegalai2019}
K. Manimegalai, C. F. S. Zephania, P. K. Bera, P. Bera, S. K. Das, T. Sil,  Study of strongly nonlinear oscillators using the Aboodh transform and the homotopy perturbation method,  Eup. Phys. J. Plus {\bf 134}, 462 (2019).
\bibitem{mehdipour2010application}
I Mehdipour, D D Ganji, and M Mozaffari, Application of the energy balance method to nonlinear vibrating equations,  Curr. Appl. Phys. {\bf 10}, 104 (2010).
\bibitem{nofal2013analytical}
T A Nofal, G M Ismail, A A M Mady, and S Abdel-Khalek, Analytical and Approximate Solutions to the Fee Vibration of Strongly Nonlinear Oscillators, J. Electromagn. Anal. Appl.  {\bf 5(10)}, 388 (2013)
\bibitem{bonham1966use}
R A Bonham, L S Su, Use of Hellmann—Feynman and Hypervirial Theorems to Obtain Anharmonic Vibration—Rotation Expectation Values and Their Application to Gas Diffraction, J. Chem. Phys. \textbf{45}, 2827 (1966)
\bibitem{bender1969anharmonic}
C M Bender, T T Wu, Anharmonic oscillator, Phys. Rev. \textbf{184}, 1231 (1969)
\bibitem{chang1975quantum}
S -J Chang, Quantum fluctuations in a $\phi^4$ field theory. I. Stability of the vacuum, Phys. Rev. D \textbf{12}, 1071 (1975)
\bibitem{hsue1984cs}
C S Hsue, J L Chern, Two-step approach to one-dimensional anharmonic oscillators, Phys. Rev. D \textbf{29}, 643 (1984)
\bibitem{ishmukhamedov2017tunneling}
I S Ishmukhamedov, V S Melezhika, Tunneling of two bosonic atoms from a one-dimensional anharmonic trap, Phys. Rev. A \textbf{95}, 062701 (2017)
\bibitem{prentice2017first}
J C Prentice, B Monserrat, R J Needs, First-principles study of the dynamic Jahn-Teller distortion of the neutral vacancy in diamond, Phys. Rev. B \textbf{95}, 014108 (2017)
\bibitem{nayfehnonlinear}
A H Nayfeh, D Mook, Nonlinear oscillations, (John Willey and Sons, New York, 1979)
%\bibitem{bogoli͡ubov1961asymptotic}
%N Bogoli͡ubov, Asymptotic methods in the theory of non-linear oscillations, 10, CRC Press (1961)
\bibitem{agrwal1985weighted}
V Agrwal, H Denman, Weighted linearization technique for period approximation in large amplitude non-linear oscillations, J. Sound Vib. \textbf{99}, 463 (1985)
\bibitem{chen1991perturbation}
S Chen, Y Cheung, S Lau, On perturbation procedure for limit cycle analysis, Int. J. Nonlin. Mech. \textbf{26}, 125 (1991)
\bibitem{cheung1991modified}
Y Cheung, S Chen, S Lau, A modified Lindstedt-Poincaré method for certain strongly non-linear oscillators, Int. J. Nonlin. Mech. \textbf{26}, 367 (1991)
\bibitem{adomian1988review}
G Adomian, A review of the decomposition method in applied mathematics, J Math. Anal. Appl. \textbf{135}, 501 (1988)
%\bibitem{nikiforov1988special}
%A F Nikiforov, V B Uvarov, Special functions of mathematical physics. A unified introduction with applications. Translated from the Russian and with a preface by Ralph P. Boas. With a foreword by AA Samarskii, (Birkh{\"a}user, Basel, Boston, 1988)
%\bibitem{bera2013exact}
%P Bera, T Sil, Pramana 80, 31 (2013)
\bibitem{marinca2019}
N.Herisanu, V. Marinca, G. Madescu, F. Dragan, Dynamic Response of a Permanent Magnet Synchronous Generator to a Wind Gust, Energies \textbf{12}, 915 (2019)
\bibitem{he2019}
N. Anjum, J -H He, Laplace transform: Making the variational iteration method easier, Appl. Math. Lett. \textbf{92}, 134 (2019)
\bibitem{liao1992proposed}
S J Liao, {\it The proposed homotopy analysis technique for the solution of nonlinear problems}, Ph. D. Thesis, Shanghai Jiao Tong University Shanghai (1992).
\bibitem{liao2009theorem}
S. J. Liao, Notes on the homotopy analysis method: some definitions and theorems, Commun. Nonlin. Sci Numer. Simulat. {\bf 14}, 983 (2009)
%\bibitem{liao1998application}
%S. J. Liao, A. T. Chwang,  J. Appl. Mech. {\bf 65} 914 (1998)
\bibitem{he1999homotopy}
J -H He, Homotopy perturbation technique, Comput. Method Appl. M. \textbf{178}, 257 (1999)
\bibitem{he2000coupling}
J -H He, A coupling method of a homotopy technique and a perturbation technique for non-linear problems, Int. J. Nonlin. Mech. \textbf{35}, 37 (2000)
\bibitem{biazar2011new}
J Biazar, M Eslami, A new homotopy perturbation method for solving systems of partial differential equations, Comput. Math. Appl. \textbf{62}, 225 (2011)
\bibitem{bera2012homotopy}
P Bera, T Sil, Homotopy perturbation method in quantum mechanical problems,  Appl. Math. Comput. \textbf{219}, 3272 (2012)
\bibitem{yildirim2009homotopy}
A Y{\i}ld{\i}r{\i}m, Retraction: “Homotopy perturbation method to obtain exact special solutions with solitary patterns for Boussinesq-like B(m,n) equations with fully nonlinear dispersion, J. Math. Phys. \textbf{50}, 023510 (2009)
\bibitem{biazar2015}
Z Ayati, J Biazar, On the convergence of Homotopy perturbation method, J. Egypt. Math. Soc. \textbf{23}, 424 (2015)
\bibitem{he2004}
J.-H He, Comparison of homotopy perturbation method and homotopy analysis method, Appl. Math. Comput {\bf 156}, 527 (2004) 
\bibitem{liao2005}
S. J. Liao,  An analytic approach to solve multiple solutions of a strongly nonlinear problem, Appl. Math. Comput. {\bf 169}, 854 (2005)
\bibitem{arfken}
G. B. Arfken, H. J. Weber, F. E. Harris, {\it Mathematical methods for physicists} (Academic Press, New Delhi, 2013)
\bibitem{hermann2014}
M Hermann, M Saravi, H. E. Khah, Analytical study of nonlinear oscillatory systems using the Hamiltonian approach technique, J Theor. Appl. Phys. {\bf 8}, 133 (2014) 
%%%% corrected
\end{thebibliography}
\end{document}